\documentclass[a4paper,12pt]{article}
\usepackage[utf8x]{inputenc}
\usepackage[T1]{fontenc}
\usepackage[colorlinks=true,linkcolor=black,citecolor=black,urlcolor=blue]{hyperref}
\usepackage{geometry}
\usepackage{titlesec}
\titleformat{\section}
{\bfseries\uppercase}{\thesection.}{1em}{}
\titleformat{\subsection}
{\bfseries}{\thesection.\thesubsection.}{1em}{}

\usepackage{graphicx} 
\usepackage{multirow} 
\usepackage{cite}
\usepackage{breakurl}
\usepackage{indentfirst}
\usepackage{amsmath, amssymb, amsfonts, bm}
\usepackage{txfonts}
\usepackage{enumitem}
\usepackage{xcolor}
\usepackage{enumitem}
\usepackage{tabularx}
\usepackage{tabularray}
\usepackage{geometry}
\usepackage{ctable}

\titlespacing*{\subsection}{0pt}{1.5em}{0.2em}
\titlespacing*{\section}{0pt}{1.5em}{0.2em}

\renewcommand\eqref[1]{Equation~\ref{#1}}

\renewcommand{\thesection}{\arabic{section}}
\renewcommand{\thesubsection}{\arabic{subsection}}

\makeatletter
\renewcommand\@biblabel[1]{#1.}
\makeatother

\newlength{\bibitemsep}
\newlength{\bibparskip}
\let\oldthebibliography\thebibliography
\renewcommand\thebibliography[1]{%
  \oldthebibliography{#1}%
  \setlength{\parskip}{\bibitemsep}%
  \setlength{\itemsep}{\bibparskip}%
}

\newcommand{\YearConf}{2024}

\newcommand{\LogoConf}{logo_IN24.jpg}
\newcommand{\CopyrightConf}{Permission is granted for the reproduction of a fractional part of this paper published in the Proceedings of INTER-NOISE \YearConf ~ \underline{provided permission is obtained} from the author(s) \underline{and credit is given} to the author(s) and these proceedings.}

\usepackage{fancyhdr}
  \fancypagestyle{firststyle}
{   \fancyhf{}
   \fancyfoot[C]{\scriptsize \CopyrightConf}
}
\begin{document}
\thispagestyle{firststyle}

\begin{center}
	\includegraphics[width=2in]{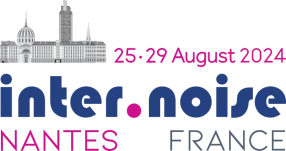}
\end{center}
\vskip.5cm

\begin{flushleft}
\fontsize{16}{20}\selectfont\bfseries

\color{black}A Survey of Integrating Wireless Technology into Active Noise Control
\end{flushleft}
\vskip1cm

\renewcommand\baselinestretch{1}
\begin{flushleft}

Xiaoyi Shen\footnote{xiaoyi003@e.ntu.edu.sg}, Dongyuan Shi, Zhengding Luo, Junwei Ji, Woon-Seng Gan\\
Nanyang Technological University\\
S2-B4A-03, 20 Nanyang Link, Singapore\\

\end{flushleft}
\textbf{\centerline{ABSTRACT}}\\
\textit{Active Noise Control (ANC) is a widely adopted technology for reducing environmental noise across various scenarios. This paper focuses on enhancing noise reduction performance, particularly through the refinement of signal quality fed into ANC systems. We discuss the main wireless technique integrated into the ANC system equipped with some innovative algorithms in diverse environments. Instead of using microphone arrays, which increase the computation complexity of the ANC system, to isolate multiple noise sources to improve noise reduction performance, the application of the wireless technique avoids extra computation demand. Wireless transmissions of reference, error, and control signals are also applied to improve the convergence performance of the ANC system. Furthermore,  this paper lists some wireless ANC applications, such as earbuds, headphones, windows, and headrests, underscoring their adaptability and efficiency in various settings. 
}
\section{INTRODUCTION}
\noindent

Active Noise Control (ANC) has become a widespread technology in attenuating noise~\cite{kuo1996active,elliot1994active,hansen1999understanding,kajikawa_gan_kuo_2012,QIU2002467,lam2021ten,shi2020active,shi2019practical, zhang2018active, zhang2021spatial,wang2024design}. ANC operates on the principle whereby a control signal is played by a secondary source that intersects with ambient noise at some specific locations, resulting in noise cancellation.  A conventional feedforward ANC system comprises reference microphones, error microphones, and secondary sources, capturing the reference and error signals and emitting the control signal, respectively~\cite{zhang2023deep, zhuang2021constrained,bai2000active, samarasinghe2016recent,wang2023experimental,shi2023active}. The control signals are generated by processing the reference signals through the adaptive control filters. To optimize the performance of the control filter, an adaptive algorithm, such as the Filtered-X Least Mean Square (FXLMS), is employed to update the weight of the control filter by minimizing a predefined cost function~\cite{lee2021review, ji2023computation,kwon2013interior,lam2020active,SHI2019651,xiao2024refracto,shen2022adaptive}.  The refinement of reference signal quality is deemed essential for the effectiveness of adaptive algorithms in ANC systems ~\cite{maeno2019spherical}. Various methodologies have been proposed to enhance the quality of the reference signals, including the utilization of multiple reference microphones in a multi-reference approach to provide comprehensive acoustic information~\cite{cheer2019application, SHEN2021107712,shen2023momentum}. Additionally, microphone arrays have been deployed to collect multiple reference signals in multi-source environments~\cite{zheng2004experimental,liu2018feed,liu2020active}. Techniques such as noise separation with microphone arrays have been employed to provide separated reference signals to the control filter ~\cite{kinoshita2015multi,iwai2019multichannel,hu2013directional,sun2020realistic}. The application of arrival time differences assists in selecting the optimal reference signal~\cite{hase2015multi}, while a time division multiple reference strategy allocates individual time slots to each reference microphone, enhancing acoustic information acquisition~\cite{ho2021time}. Moreover, the selection of dominant components in reference signals through the singular value decomposition of the power spectrum matrix~\cite{akiho1999practical} and the reconstruction of reference signals via noise signal decomposition and multi-network reconstruction to provide accurate information for the control filter~\cite{yang2022adaptive}, alongside spectrum reshaping for enhancing the correlation between the reference signal and disturbance ~\cite{chen2015improving}, have been explored.

Innovations in wireless technology present another method for enhancing signal quality. Various studies have investigated the integration of wireless transmission with ANC, leveraging the speed of wireless transmission over sound to afford ANC controllers advanced signal reception and processing time. Strategies include placing wireless transmitters at the door of the room to capture incoming reference signals ~\cite{shen2018mute,janveja2023winc}, and positioning transmitters close  to noise sources~\cite{shen2024principle,shen2022hybrid,shen2023implementations}. Alternatively, wireless technology has been utilized to transmit error signals~\cite{shi2020simultaneous} and control signals processed by adaptive filters~\cite{luo2022ec}. Notably, wireless feedforward ANC headphones employ coherence-based signal selection, and coherence-based weight determination methods to optimize noise reduction~\cite{ 
 shen2022multiv}, and wireless hybrid ANC systems utilize an error separation module to individually update feedforward and feedback filters with separated error signals~\cite{shen2022hybrid}, and apply the fixed-adaptive selection control for distinct feedforward and feedback filter updates.  Wireless earbuds, incorporating microphones, controllers, and wireless modules within their charging cases, transmit control signals wirelessly to the earbuds~\cite{luo2022ec}. Similarly, wireless ANC headrests and windows apply wireless reference signals to enhance the noise reduction capabilities of multi-channel ANC systems~\cite{shen2023implementations, shen2024principle}. Moreover, the incorporation of wireless ANC technology in infant incubators highlights its broader applicability~\cite{liu2016wireless}.

This paper provides a comprehensive overview of adaptive algorithms employed in wireless ANC technologies. Multiple algorithms applied in wireless ANC methodologies are discussed in Section~\ref{sec_2}. Section~\ref{sec_4} introduces various applications of wireless ANC, illustrating the broad usage of wireless ANC.  Section~\ref{sec_5} concludes the whole paper.

\section{Wireless technique integrated with ANC} \label{sec_2}
\noindent
This section introduces various ANC algorithms equipped with wireless technology. The time-domain/frequency-domain lookahead aware algorithm, digital twin technology, coherence-based selection, and coherence-based weight determination are applied in the wireless feedforward ANC. Error separation module and fixed-adaptive control selection technique are implemented in the wireless hybrid ANC.

\subsection{Filtered-X least mean square algorithm }
The primary goal of ANC is to generate the control signal played by the secondary source to cancel the noise. The control signal $y(n)$ is generated by the reference signal $\bm{x}(n)$ passing through the control filter
\begin{equation}\label{eq_30}
y(n) = \mathbf{x}^\mathrm{T}(n)\mathbf{w}(n),
\end{equation}
where $\mathbf{w}(n)$ denotes the coefficients of the control filter, and the reference vector $ \mathbf{x}(n)$ is given by
\begin{equation}
    \mathbf{x}(n) = \begin{bmatrix}
    x(n) & x(n-1) & \cdots & x(n-L+1)
    \end{bmatrix}^\mathrm{T},
\end{equation}
in which $L$ denotes the length of the control filter. 

The error signal $e(n)$ received at the error microphone side is represented as
\begin{equation}
    e(n) = d(n)-y(n) \ast s(n),
\end{equation}
where $d(n)$ represents the disturbance arrived at the error microphone,  and $s(n)$ stands for the impulse response of the secondary path from the secondary source to the error microphone. 

In order to minimize the error signal, FXLMS algorithm is applied to update the coefficients of the control filter
\begin{equation}\label{weight_update2}
    \mathbf{w}(n+1) =\mathbf{w}(n)+\mu \mathbf{x}'(n)e(n), 
\end{equation}
where $\mu$ denotes the step size, and the filtered reference signal is given by
\begin{equation}
     \mathbf{x}'(n) = \mathbf{x}(n) \ast \hat{s}(n), 
\end{equation}
where $\hat{s}(n)$ is the secondary path's estimated impulse response, which can be measured online or offline.

The FXLMS algorithm is commonly used in ANC to update the coefficients of the control filter. According to the updating equation, the reference signal is crucial for the effectiveness of the FXLMS algorithm. Consequently, wireless lookahead-aware ANC (WLANC) attempts to utilize wireless transmission to deliver enhanced information from the reference signal to the control filter.

\subsection{Wireless lookahead aware ANC (WLANC)}
In the implementation of WLANC, the wireless microphone is placed at the door of the room to receive the reference signal and transmit it to the control filter faster than the sound traveling through the air. To utilize this future sample of the reference signal, the wireless Lookahead aware ANC (WLANC) model~\cite{shen2018mute} is applied to update the control filter
\begin{equation}\label{weight_update}
    w_k(n+1)=w_k(n)+\mu \mathbf{x}'(n+k)e(n), 
\end{equation}
where $w_k(n)$ denotes the $k$th coefficient of the control filter ($k=1,2,\cdots,L$) and  is updated by the look-ahead reference signal transmitted by the wireless microphone and the control signal is generated by:
\begin{equation}
    y(n) = \sum_{k=N}^{L+N} w_k(n)x(n+k),
\end{equation}
where $N$ denotes the lookahead samples brought by wireless transmission.  This significant lookahead capability enhances the noise reduction efficacy of the ANC system.

The convergence speed is degraded in a multi-source environment with $J$ reference microphones.  To overcome this challenge, a frequency-domain adaptive filtering solution is applied~\cite{janveja2023winc}. DFT is applied to convert time-domain convolutions into frequency-domain multiplications. In the frequency domain, the coefficient of the adaptive filter is calculated in parallel in each frequency bin
\begin{equation}
   W_j(f) \leftarrow W_j(f) + \mu_j(f) E^\mathbb{T}(f) X_j(f) S(f), j=1,2,\cdots,J.
\end{equation}
where $\mu_j(f)$ is the adaptive step size for $j^{th}$ control filter
\begin{equation}
    \mu_j(f) = \frac{\mu_j}{P_j(f)}.
\end{equation}
Here, $\mu_j$ is a constant, and $P_j(f)$ is the measure of mean power in this specific frequency bin of the reference signal. Selecting different step sizes for different frequency bins also results in faster convergence speeds for signals with lower power.

\subsection{Wireelss networked adaptive ANC with digital-twin FXLMS algorithm (WANC-DT)}
\begin{figure}
    \centering
    \includegraphics[width = 0.9 \textwidth]{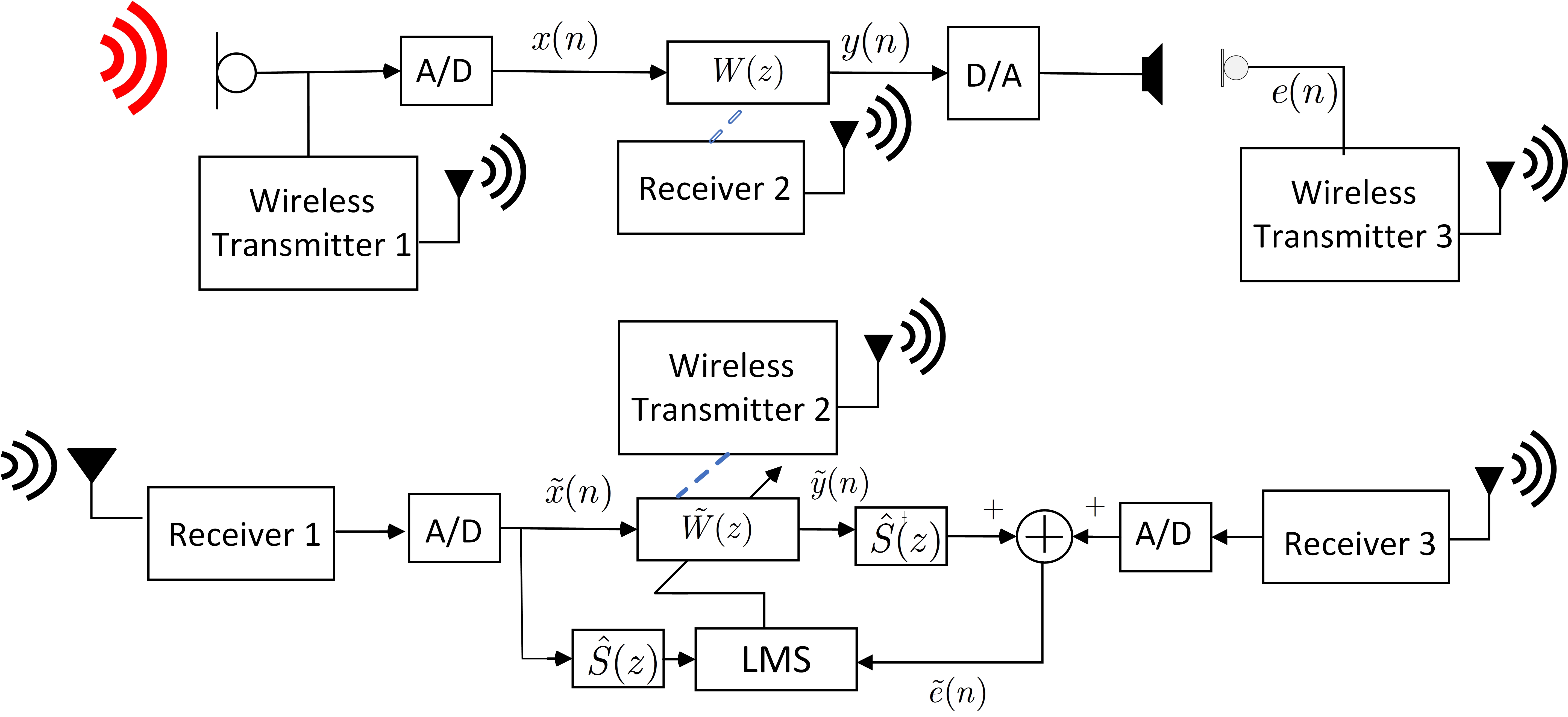}
    \caption{The block diagram of wireless networked adaptive ANC with digital-twin FXLMS algorithm}
    \label{fig_DIGITALTWIN}
\end{figure}

Another application of the wireless technique to ANC is the wireless networked adaptive ANC with digital-twin FXLMS algorithm (WANC-DT).  The wireless transmission collaborates with cloud computing technologies for the ANC system~\cite{shi2022digital}. The reference and error signals are transmitted wirelessly to the cloud, where the control filter is updated on the cloud side (referred to as the digital twin) using the FxLMS algorithm. The updated coefficients are then sent back wirelessly to the local site to produce the control signal, as illustrated in Fig.\ref{fig_DIGITALTWIN}.
In the digital twin, the reference signal is transmitted wirelessly from the local side
\begin{equation}
    \tilde{\mathbf{x}}(n) = \left[ x(n),x(n-1),\cdots,x(n-L+1)     \right],
\end{equation}
and the error signal is calculated as
\begin{equation}
    \tilde{e}(n) = e(n) + \hat{s}(n) \ast \tilde{y}^\mathrm{T}(n).
\end{equation}
The incremental vector, which contains the control filter coefficients, is downloaded to the local controller. Consequently, the local controller updates its control filter by
\begin{equation}
    \mathbf{w}(n'+1) =  \mathbf{w}(n') + \tilde{\mathbf{w}}(n+1),
\end{equation}
where $n'$ is the current time index in the local controller and $\tilde{\mathbf{w}}(n+1) $ denotes the coefficients updated in the digital twin.

The digital twin (DT)-FxLMS algorithm simplifies the local controller's tasks by eliminating the need to compute the filtered reference signal and update the control filter coefficients. Instead, the local controller focuses solely on generating the control signal in real-time and transmitting the sensing data to the cloud. Consequently, the computational burden of the DT-FxLMS algorithm on the local controller is nearly equivalent to that of employing a pre-trained fixed filter. However, since these methods transmit the reference, error, and control signals during the ANC process, the synchronization of the signal should be considered.

\subsection{Wireless ANC with coherence-based-selection technique (WANC-CBS)}
While the two above-mentioned algorithms wirelessly transmit the reference signal to the controller, they still capture the reference signal with a low RSIR. In a multi-source environment, the RSIR significantly impacts the noise reduction performance of the ANC system. The reference signal $x(n)$ of the ANC system is assumed to interfere with uncorrelated noise $v(n)$. RSIR is calculated as:
\begin{equation}
    \text{RSIR}=\frac{\mathbb{E}\left[x^2(n)\right]}{\mathbb{E}\left[v^2(n)\right]}=\frac{\sigma^2_{x}}{\sigma^2_v},
\end{equation}
where  $\mathbb{E}[\cdot]$, $\sigma^2_{x}$, and $\sigma^2_v$ denote the expectation operator, and the variances of $x(n)$ and $v(n)$. The noise reduction performance is influenced by RSIR

\begin{equation}\label{MMSE_eq41}
    \text{NR} \le 10\log_{10}\left(\frac{L\sigma^2_x}{\sigma^2_v}\right)=10\log_{10}(L\cdot\text{RSIR}).    
\end{equation}
 A low RSIR can adversely affect the ANC system's ability to reduce noise~\cite{shen2024principle}. Employing wireless technology to capture the reference signal close to noise source can markedly enhance the RSIR, thereby improving the noise reduction performance of the ANC system.

\begin{figure}
    \centering
    \includegraphics[width =0.9 \textwidth]{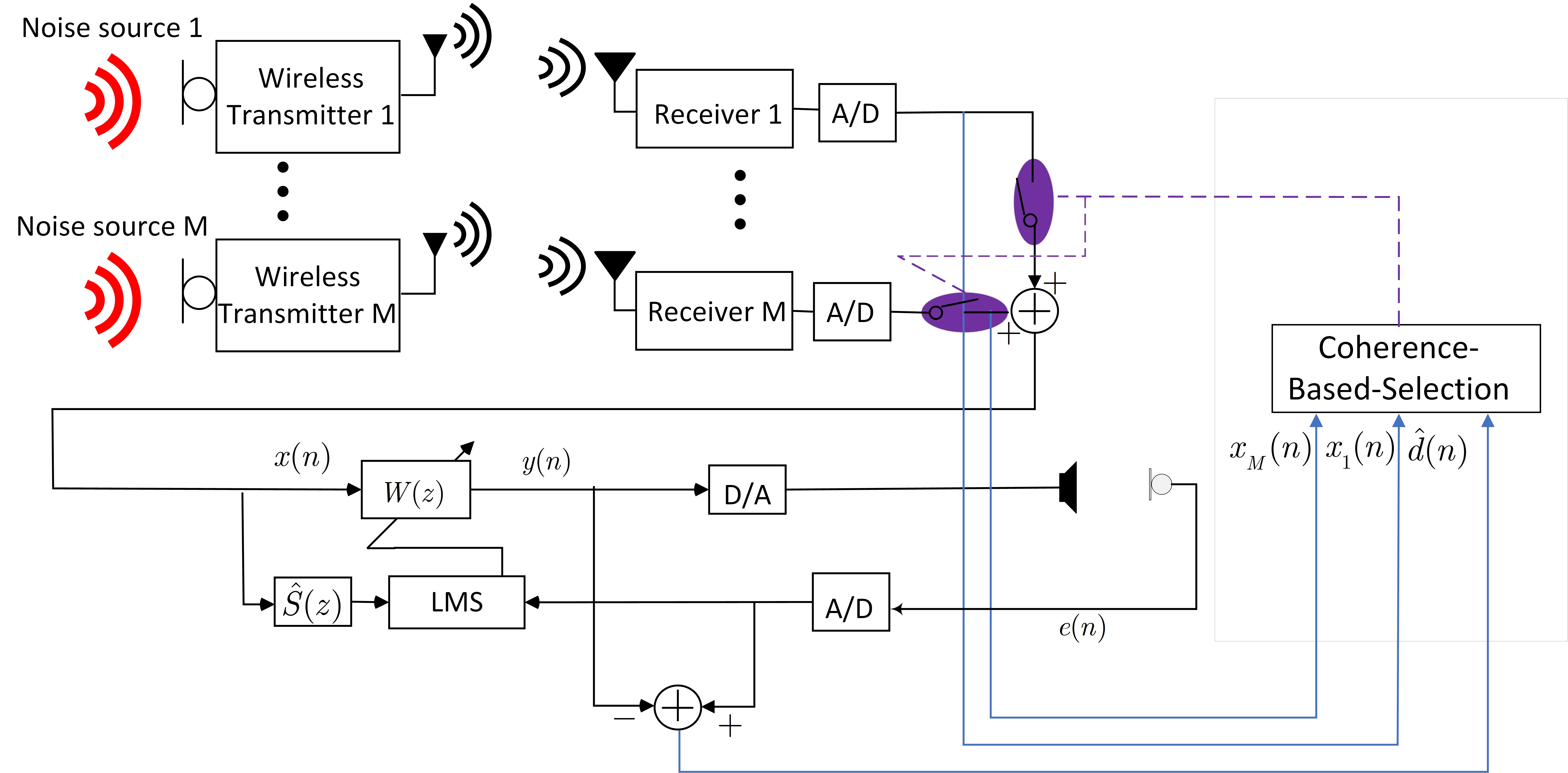}
    \caption{The block diagram of wireless ANC implemented with coherence-based selection method.}
    \label{fig_cbs}
\end{figure}

To identify the reference signal received by the wireless microphone that contributes to the disturbance, a coherence-based selection method~\cite{shen2021implementation,9414683} is applied in wireless ANC, illustrated in Fig.\ref{fig_cbs}.
In a multi-source environment where $M$ individual noise sources exist in the surroundings
\begin{equation}
  \mathbf{x}(n)=\left[x_1(n),x_2(n), \cdots, x_M(n)\right].  
\end{equation}
The magnitude squared coherence (MSC) between $m$th reference signal collected by the wireless microphone and the estimated disturbance is calculated as
\begin{equation}
   C_{x_m\hat{d}}(\omega)= \frac{1}{S_{\hat{d}\hat{d}}(\omega)} S_{x_m\hat{d}}^{*}(\omega)S_{x_mx_m}^{-1}(\omega)S_{x_m\hat{d}}(\omega),
\end{equation}
where $m=1,2,\cdots,M$, and  $\omega$ denotes the frequency of interest. $S_{x_m\hat{d}}(\omega)$ represents the cross-power spectrum of the estimated disturbance $\hat{d}(n)$ and the reference signal $x_m(n)$. $S_{x_mx_m}(\omega)$ and $S_{\hat{d}\hat{d}}(\omega)$ denote the auto-power spectrum of  $x_m(n)$ and $d(n)$, respectively. The estimated disturbance $\hat{d}(n)$ is calculated as
\begin{equation}\label{equation2}
  \hat{d}(n)=e(n) + y(n)\ast \hat{s}(n).
\end{equation}
 To select the reference signals that contribute to the disturbance and discard the ineffective reference signals, a threshold value of $C_T$ is used
\begin{equation}\label{eq_5_9}
 C_{x_m\hat{d}} \ge C_T.
\end{equation}
Since the noise reduction performance is related to MSC, the threshold $C_T$ can be determined by
\begin{equation}\label{eq_5_10}
 C_T =1-10^{(-\text{NR}/10)},
\end{equation}
where $\text{NR}$ is the expected noise reduction at the error microphone side.

With the selection of the high-coherent reference signal, the wireless ANC system eliminates the need for unnecessary computations for low-coherent reference signals. Rather than just selecting or discarding the reference, an enhanced method that applies weight to the reference which is derived from MSC.
\subsection{Wireless ANC with coherence-based weight determination technique (WANC-CWD)}
\begin{figure}
    \centering
    \includegraphics[width =0.9 \textwidth]{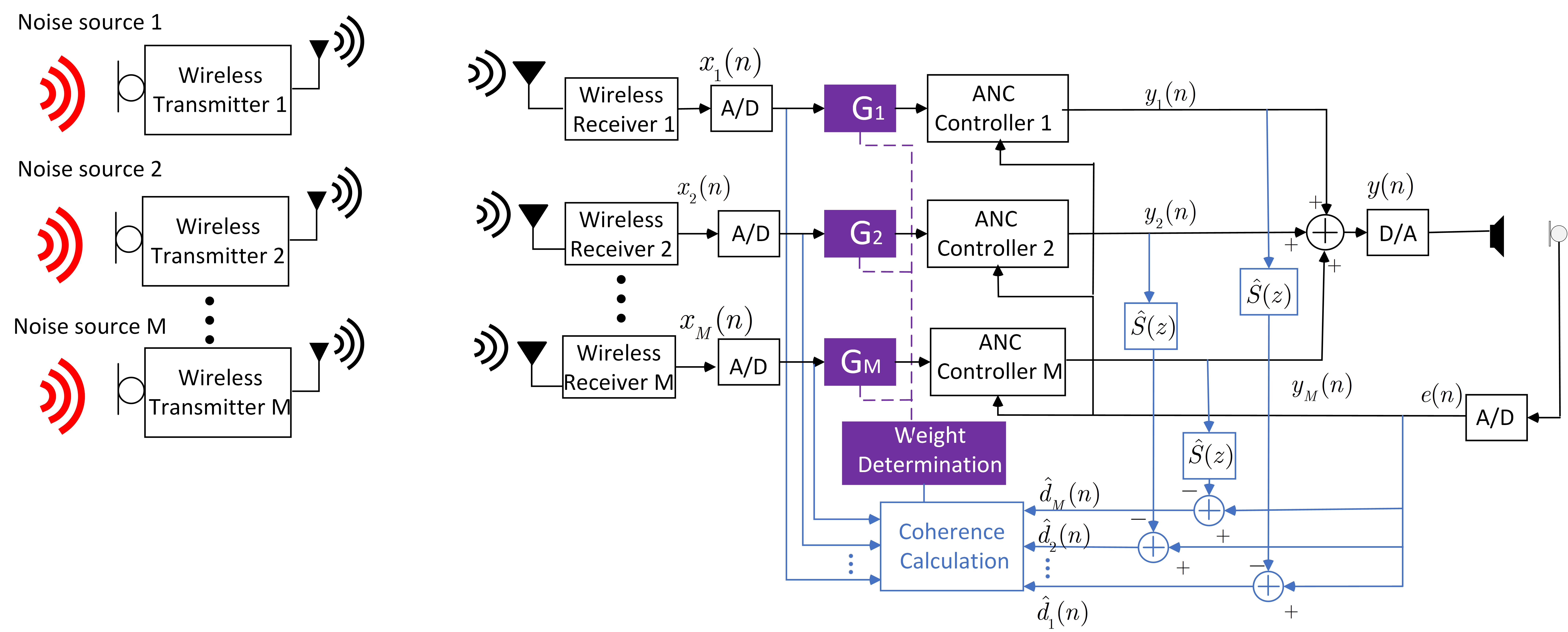}
    \caption{The block diagram of wireless ANC implemented with coherence-based weight determination method.}
    \label{fig_cwd}
\end{figure}

To maximize the potential of the wireless reference signal,  a coherence-based weight determination (CWD) technique \cite{shen2022multiv} is proposed. This approach involves assigning weights to various references instead of merely selecting or discarding them. The block diagram depicting the wireless ANC incorporating CWD is illustrated in Fig.~\ref{fig_cwd}. 
 Instead of CBS, CWD calculates the weight according to MSC
\begin{equation}\label{eq_5_18}
     G_m =  \begin{cases} 
                0,  & C_m<C_{T1}, \\
                \frac{C_m-C_{T1}}{C_{T2}-C_{T1}}, & C_{T1}<C_m<C_{T2},\\
                1.   &C_m>C_{T2}.
            \end{cases}  m=1,2,\cdots, M
\end{equation}
where $C_{T1}$ and $C_{T2}$ is determined by
\begin{equation}\label{eq_5_19}
    \begin{cases}
         C_{T1} =1-10^{(-\text{NR}_1/10)},\\
         C_{T2} =1-10^{(-\text{NR}_2/10)},
    \end{cases}
\end{equation}
where $\text{NR}_1$ and $\text{NR}_2$ represent the minimum and maximum noise reduction levels aimed to achieve at the error microphone.
The received $m$th reference signal $x_m(n)$ is filtered by the $m$th control filter $W_m(z)$, and the output signal $y_m(n)$ is given by
\begin{equation}\label{eq_5_20}
  y_m(n)=G_m\mathbf{x}_m^\mathrm{T} (n)\mathbf{w}_m(n), 
\end{equation} 
where $\mathbf{w}_m(n)$ is the weight vector of the $m$th control filter. 

Both wireless ANC with the CBS and CWD receive the reference signal with a high RSIR, enhancing the noise reduction performance. However, these systems struggle to attenuate the unexpected noises that the wireless microphone does not pick up. Therefore, we introduce a hybrid wireless structure in the following section.

\subsection{Wireless hybrid ANC with error separation module (WHANC-ESM)}
\begin{figure}
    \centering
    \includegraphics[width =0.9 \textwidth]{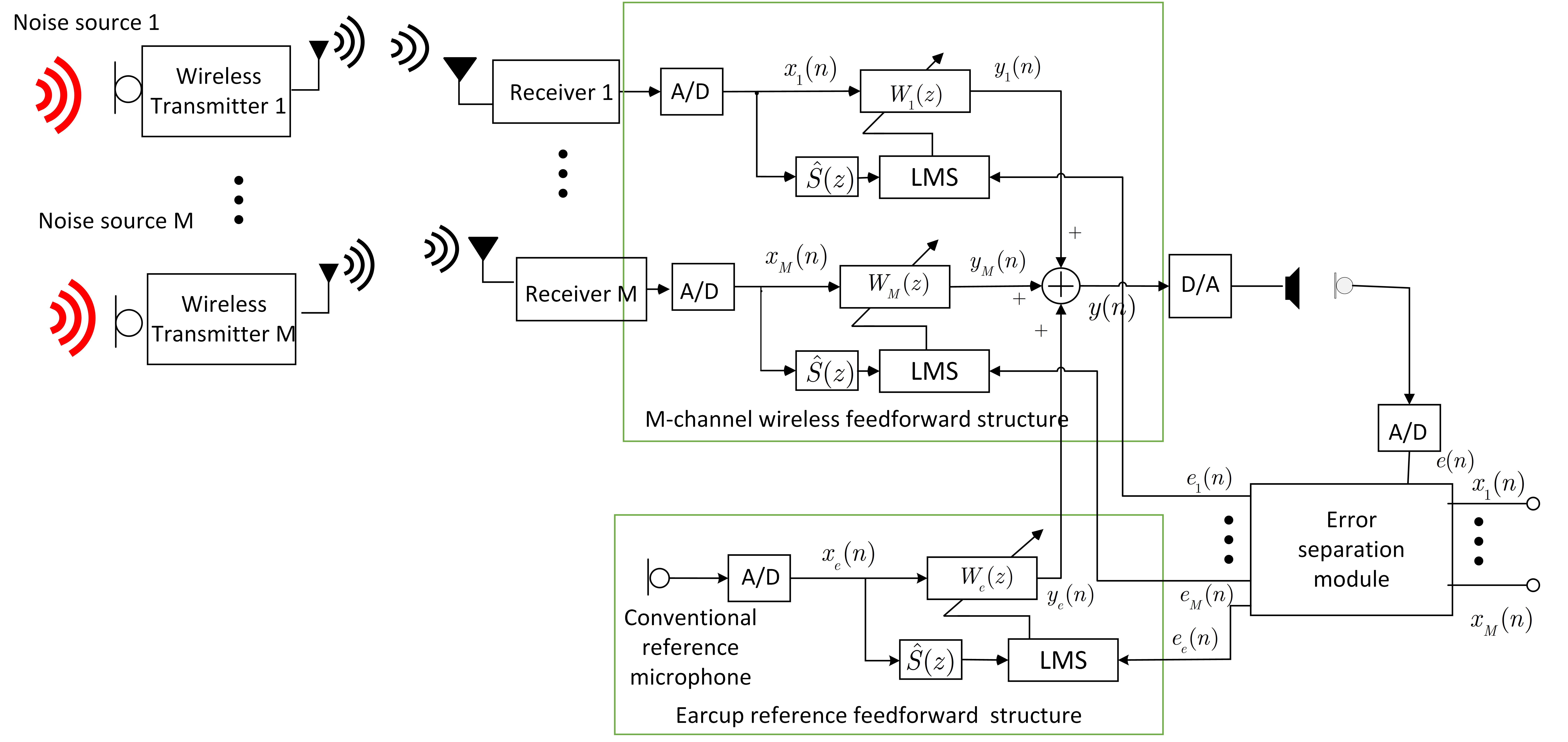}
    \caption{The block diagram of wireless hybrid ANC implemented with error separation module.}
    \label{fig_earcup}
\end{figure}

Given that a wireless microphone can only attenuate noise that has been previously identified, it fails to deal with the noise that occurs unexpectedly in the environment. To overcome this limitation and improve noise reduction performance, a hybrid approach combining wireless and conventional microphones is utilized. This strategy not only attenuates uncorrelated noise but also helps in improving overall noise reduction performance ~\cite{shen2022hybrid}, as depicted in Fig.\ref{fig_earcup}.

The control signal is generated by the wireless reference signal $\mathbf{x}_m(n)$ and  the conventional reference signal $\mathbf{x}_e(n)$
\begin{equation}\label{eq_6_19}
y(n) =   \sum_{m=1}^M y_m (n) + y_e (n),
\end{equation}
where $y_e(n)$ and $y_m (n)$ represent the output of $W_e(z)$ and $W_m(z)$, separately
\begin{equation}\label{eq_6_20}
\begin{cases} 
y_m (n) = \mathbf{x}^\mathrm{T}_m(n)\mathbf{w}_m (n), ~~m=1,2,\cdots,M \\
y_e (n) = \mathbf{x}^\mathrm{T}_e(n) \mathbf{w}_e (n).
\end{cases}
\end{equation}
The error separation module (ESM) employs the separated error signal to update the adaptive filters $W_m(z)$ for the wireless reference signal and $W_e(z)$ for the conventional reference signal, respectively. The separated error signal for the $m$th channel is generated using the filter $H_m(z)$, as
\begin{equation}\label{eq_6_23}
  e_m(n) = \mathbf{x}^\mathrm{T}_m(n)\mathbf{h}_m (n), ~~m=1,2,\cdots,M .
\end{equation}
where $\mathbf{h}_m(n)=\left[h_{m1} (n),\cdots,h_{mL} (n)\right]^\mathrm{T}$ represents the coefficients of the $m$th adaptive filter. The error signal used to update $W_e(z)$ is obtained from
\begin{equation}\label{eq_6_24}
   e_e(n) = e(n) - \sum_{m=1}^M e_m(n).
\end{equation} 
and the coefficients of $H_m(z)$ is updated with step size $\mu_h$ 
\begin{equation}\label{eq_6_25}
  \mathbf{h}_m(n+1) = \mathbf{h}_m(n) + \mu_h \mathbf{x}_m(n)e_{am}(n),
\end{equation}
where $e_{am}(n)$ is the subtraction of the measured error signal $e(n)$ and the output of the filter $H_m(z)$.
In this hybrid structure, the wireless microphone captures the identified reference signal, while a conventional microphone is used to detect unexpected environmental noise. By integrating these two components, the hybrid wireless ANC system effectively attenuates unexpected noise.

\subsection{Wireless hybrid ANC with fixed-adaptive control selection technique (WHANC-FAS) }
\begin{figure}
    \centering
    \includegraphics[width =0.9 \textwidth]{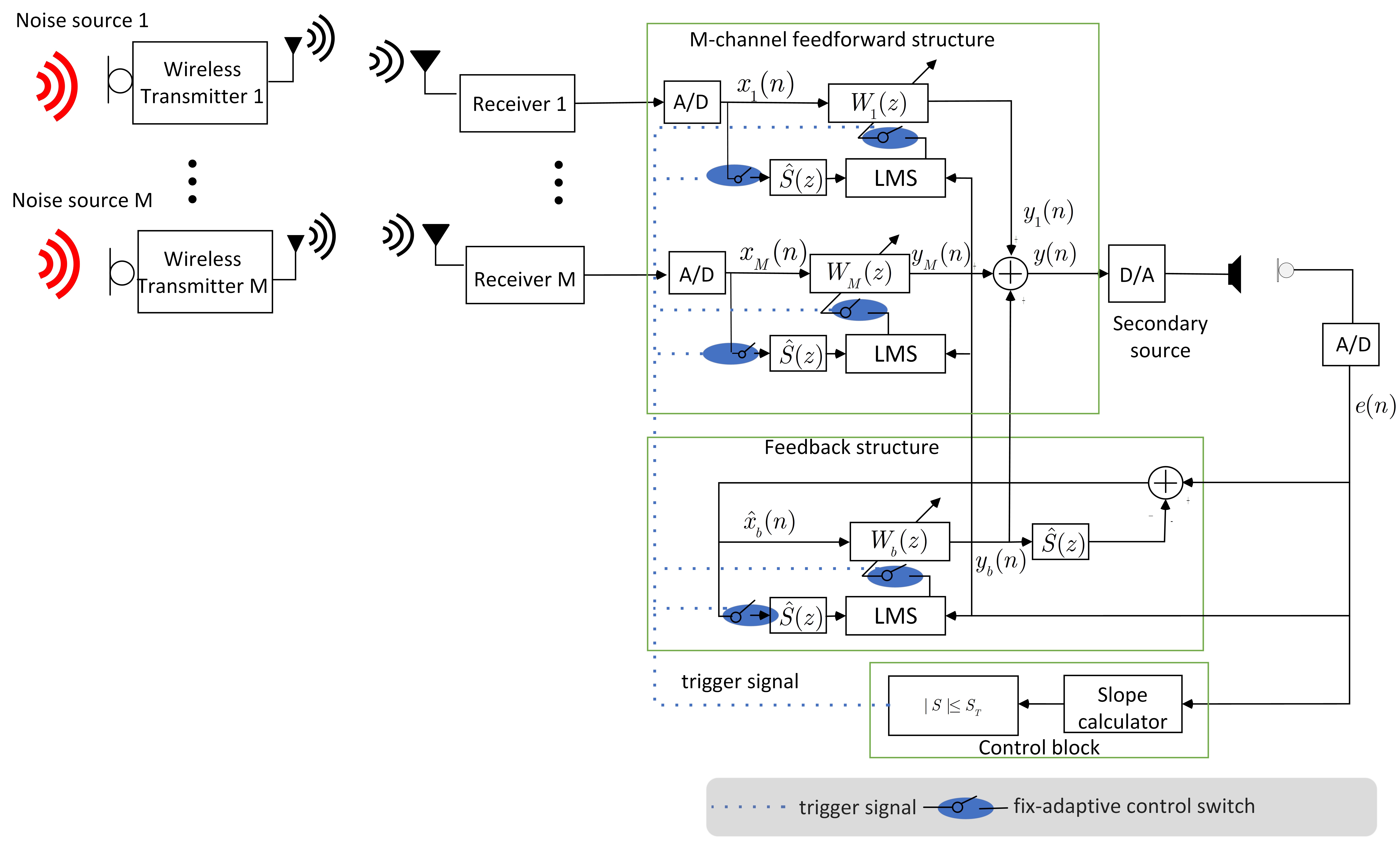}
    \caption{The block diagram of wireless hybrid ANC implemented with fixed-adaptive control selection.}
    \label{fig_esm}
\end{figure}

An alternative approach to attenuating the uncorrelated noise is using a combination of feedforward-feedback structure, as shown in Fig.\ref{fig_esm}.
The feedforward ANC control filter deals with the wireless reference signal, while the feedback control filter attenuates uncorrelated noise. The output of the control filters $y_m(n)$ and  $y_b(n)$ are calculated as
\begin{equation}\label{eq_6_53}
\begin{cases}
     y_m(n)=\mathbf{x}_m^\mathrm{T} (n)\mathbf{w}_m(n), ~~~~m=1,2,\cdots,M. \\
    y_b(n) = \mathbf{\hat{x}}^\mathrm{T}_b(n) \mathbf{w}_b(n),
\end{cases}
\end{equation} 
where $\mathbf{w}_m(n)$ and $\mathbf{w}_b(n)$ denote the coefficient of feedforward control filter $W_m(z)$ and feedback control filter $W_b(z)$. The estimated input for the feedback control filter is calculated as
\begin{equation}\label{eq_6_55}
    \hat{x}_b(n) = e(n) + y_b(n) \ast \hat{s}(n), 
\end{equation}
To reduce the computation complexicity, a fixed-adaptive control selection (FAS) is applied and divides the whole ANC process into three states \cite{shen2022jsv}
\begin{itemize}
    \item State 1: Training Mode:
    \begin{equation}\label{eq_6_10}
\begin{cases}
\mathbf{w}_\mathrm{m}(n+1)=\mathbf{w}_\mathrm{m}(n)-\mu_\mathrm{f}\mathbf{x}^\prime_\mathrm{m}(n)e(n),  \\
\mathbf{w}_\mathrm{b}(n+1) = \mathbf{0}.
  \end{cases}
\end{equation}
    \item State 2: Steady-state Mode:
    \begin{equation}\label{eq_6_13}
\begin{cases}
\mathbf{w}_\mathrm{m}(n+1)=\mathbf{w}_\mathrm{m}(n),~~~m=1,2,\cdots, M. \\
  \mathbf{w}_\mathrm{b}(n+1)=\mathbf{w}_\mathrm{b}(n).
  \end{cases}
\end{equation}
    \item State 3: (Near Field) Interference Mode: 
    \begin{equation}\label{eq_6_16}
\begin{cases}
\mathbf{w}_\mathrm{b}(n+1)=\mathbf{w}_\mathrm{b}(n)-\mu_\mathrm{b}\mathbf{x}^\prime_\mathrm{b}(n)e(n),
  \\\mathbf{w}_\mathrm{m}(n+1)=\mathbf{w}_\mathrm{m}(n),~~~m=1,2,\cdots, M.  
\end{cases}
\end{equation}
\end{itemize}
If there is no unexpected noise occurring in the environment, WHANC-FAS initially transits to State 1 and updates the coefficients for the $M$ feedforward control filter. Once the algorithm converges, FAS forces the state change to State 2, where the coefficients of the control filters are fixed and no further updates are made to save computation complexity. When WHANC-FAS detects the unexpected noise that is not captured by the wireless microphone, the algorithm switches to State 3, during which it begins updating the feedback control filter until convergence is achieved.

\subsection{Comparative evaluation of wireless ANC}
Multiple algorithms are applied to wireless ANC to improve the convergence performance, as detailed in Table~\ref{table_1}. WLANC uses reference signals with advanced knowledge and processes the control filter in the frequency domain to increase convergence speed. WANC-DT shifts the adaptive updating process to the cloud controller, thereby reducing the computational complexity at the local level. However, both utilize the mixed reference signals and do not exhibit improved noise reduction performance. WANC-CBS and WANC-CWD position the wireless microphone close to the noise source to capture the reference signal with a high RSIR, which enhances noise reduction performance. However, these feedforward wireless ANC systems do not effectively cancel the unexpected noises. WHANC-ESM and WHANC-FAS employ conventional microphones or feedback structures to address unexpected noise and improve the noise reduction performance of wireless ANC.
\begin{table}[!t]
\caption{Comparative evaluation of the different types of wireless ANC algorithm}\label{table_1}
\begin{tabular}{|c|l|l|}
\hline
Algorithms & \multicolumn{1}{c|}{Advantages}                                                                                                                                                                & \multicolumn{1}{c|}{Disadvantages}                                                                                                                                                                                            \\ \hline
WLANC       & \begin{tabular}[c]{@{}l@{}}$\cdot$ Applies future samples of reference \\     signals.\\ $\cdot$ Frequency-domain adaptive filtering \\     increases the convergence speed\end{tabular}                & \begin{tabular}[c]{@{}l@{}}$\cdot$ DFT increases computation burden\\ and time delay.\\ $\cdot$ Received mixed reference signals. \end{tabular}                                                                                                           \\ \hline
WANC-DT    & \begin{tabular}[c]{@{}l@{}}$\cdot$ Apply cloud controller to reduce\\ the computation complexcity of\\  local controller.\end{tabular}                                            & \begin{tabular}[c]{@{}l@{}}$\cdot$ Signal synchronization issues for \\     wireless transmission of \\ reference, error and control signal.  \\$\cdot$  Noise reduction performance \\is not improved.\end{tabular}                                                                    \\ \hline
WANC-CBS   & \begin{tabular}[c]{@{}l@{}}$\cdot$ Select the coherent wireless reference\\     signal, avoiding unnecessary \\     computation and improving the noise \\     reduction performance.\end{tabular} & \multirow{2}{*}{\begin{tabular}[c]{@{}l@{}}$\cdot$ Need to place wireless microphone \\     in front of the noise in advance.\\$\cdot$  Can't  deal with the uncorrelated noise \\ that occurred in the environment.\end{tabular}} \\ \cline{1-2}
WANC-CWD   & \begin{tabular}[c]{@{}l@{}}$\cdot$ Add a calculated weight value to the \\     wireless reference signal according to\\ coherence to improve noise reduction.\end{tabular}                         &                                                                                                                                                                                                                               \\ \hline
WHANC-ESM  & \begin{tabular}[c]{@{}l@{}}$\cdot$ Attenuate the unexpected noise.\\ $\cdot$ ESM provides the separated error \\  to improve the convergence speed.\end{tabular}                        & \begin{tabular}[c]{@{}l@{}}$\cdot$ Additional conventional microphone \\     and ESM increase the computation\\     complexcity.\end{tabular}                                                                                      \\ \hline
WHANC-FAS  & \begin{tabular}[c]{@{}l@{}}$\cdot$ Deal with unexpected noise.\\ $\cdot$ FAS determines whether to \\     adaptively update the control filter \\     to reduce the computation burden.\end{tabular}    & \begin{tabular}[c]{@{}l@{}}$\cdot$ For unexpected noise cancellation,\\     only be able to attenuate \\     uncorrelated narrowband noise.\end{tabular}                                                                           \\ \hline
\end{tabular}
\end{table}

\section{Applications of wireless ANC} \label{sec_4}
Wireless ANC technology finds extensive application across various devices. In wireless ANC headphones, a wireless receiver is utilized to capture reference signals, control signals, and error signals. For wireless ANC headrests, the wireless microphones positioned on the ground capture the reference signals for the control filter implemented inside the headrest. Meanwhile, wireless windows employ the wireless microphones outside the building to receive reference signals.

The adoption of wireless technology in ANC earbuds and headphones is widespread. An advanced ANC earbud integrates the reference microphone, controller, and wireless module within the charging case, whereas the speaker and error microphone are embedded within the earbud. Control signals are wirelessly sent from the charging case to the earbud, and error signals are sent back from the earbud to the charging case~\cite{luo2022ec}. This configuration helps to reduce the computation burden of the earbud since all the calculations are completed in the charging case. Placing the wireless transmitter in proximity to noise sources enables the delivery of reference signals to the headphone's controller~\cite{shen2022jsv}. This arrangement significantly increases the distance between the reference and error microphones ($d_e-d_r$) compared to traditional ANC headphones, providing additional lookahead time $(d_e-d_r)/v_{\text{sound}}$ for the ANC control filter to process the control signal. Another innovation involves wireless IoT-based noise cancellation, which transmits ambient noise data to headphones via wireless connections and attenuates multiple noise sources using an IoT node network. \cite{janveja2023winc}.

The wireless ANC headrest~\cite{shen2023implementations} features the wireless microphone on the ground to wirelessly send the reference signal to the control filter located behind the head. The headrest in the car, equipped with wireless transmission, can receive improved reference signals from the road and engine, thereby improving noise reduction performance. This setup includes two secondary speakers and two error microphones, providing quiet zones for both the left and right ears.

The concept of wireless ANC has been adapted for use in residential windows, enabling comprehensive noise control throughout an entire window~\cite{shen2024principle}. The ANC controller, situated near the windows, generates control signals to attenuate noises. 
The wireless microphone can be positioned anywhere to capture potential noise and transmit the reference signal to multiple ANC controllers, facilitating noise reduction throughout the entire building. Additionally, the use of wireless microphones allows ANC windows to reduce noise whether they are open or closed, as the wireless signal can penetrate glass and walls.

\section{Conclusion} \label{sec_5}
This paper has reviewed the integration of wireless technologies into ANC systems and discussed various algorithms developed for wireless ANC. Time-domain and frequency-domain lookahead-aware algorithms in wireless ANC supplied advanced reference samples to enhance convergence speed. The digital-twin FXLMS algorithm utilized both local and cloud controllers to minimize computational complexity. However, both approaches utilized mixed reference signals, which degraded noise reduction performance. For the noise source which can be identified in advance, wireless ANC with coherence-based selection and coherence-based weight determination were applied with the high reference signal-to-interference ratio signals. Additionally, to deal with unexpected noise in the environment, the error separation module and fixed-adaptive control selection were incorporated into the wireless hybrid ANC with the increased computational complexity. These advancements have expanded the use of wireless ANC in devices like earbuds, headphones, headrests, and windows. With the continuous development of IoT networks and wireless technologies, further improvements and broader applications of wireless ANC are expected.

\section*{Acknowledgements}
\noindent
This research is supported by the Singapore Ministry of Education Academic Research Fund Tier 2 (Award No.\@ MOE-T2EP50122-0018).

\bibliographystyle{unsrt}
\bibliography{a} 

\end{document}